\documentclass[12pt]{article}
\usepackage[pdfstartview=FitH,colorlinks=true,linkcolor=blue,anchorcolor=red,citecolor=magenta,urlcolor=blue]{hyperref}
\usepackage[english]{babel}
\usepackage{amsmath,amssymb,titling,authblk}
\usepackage{slashed}
\usepackage{amsmath}
\usepackage{amscd}
\usepackage[normalem]{ulem}
\usepackage{appendix}
\usepackage{bbold}
\usepackage{bm}

\usepackage{pdflscape}
\usepackage{yfonts}
\usepackage{amscd}
\usepackage{epsfig}
 \usepackage{microtype}
 
 \usepackage[utf8]{inputenc}
\usepackage[T1]{fontenc}

\usepackage{cite}

\allowdisplaybreaks[4]
\numberwithin{equation}{section}

\usepackage {color}
\usepackage{bbold}
\usepackage{braket}
\usepackage{bbm}

\definecolor{verde}{cmyk}{.83,.21,1,.08}

\advance\hoffset by -1.1truecm
\advance\voffset by -1.0truecm
\newcommand{\be}{\begin{equation}}
\newcommand{\ee}{\end{equation}}
\newcommand{\bea}{\begin{eqnarray}}
\newcommand{\eea}{\end{eqnarray}}

\usepackage{mathrsfs}
\usepackage{authblk}

\numberwithin{equation}{section}
\usepackage{float}
\restylefloat{figure}
\newcounter{appendice}

\usepackage[T1]{fontenc}
\usepackage{lmodern}

\begin{document}

\setlength{\droptitle}{-6pc}

\title{316}

\renewcommand\Affilfont{\itshape}
\setlength{\affilsep}{1.5em}

\author[1,2,3]{Fedele Lizzi\thanks{fedele.lizzi@na.infn.it, fedele.lizzi@unina.it}}
\affil[1]{Dipartimento di Fisica ``Ettore Pancini'', Universit\`{a} di Napoli {\sl Federico~II}, Napoli, Italy}
\affil[2]{INFN, Sezione di Napoli, Italy}
\affil[3]{Departament de F\'{\i}sica Qu\`antica i Astrof\'{\i}sica and Institut de C\'{\i}encies del Cosmos (ICCUB),
Universitat de Barcelona, Barcelona, Spain}

\date{To be published in the Festschrift volume: \textit{Particles, Fields and Topology: Celebrating A.P. Balachandran}, to be published by World Scientific, Singapore.}

\maketitle



%

\begin{abstract}
A room, a teacher and many friends.
\end{abstract}

\medskip



Let us begin with a celebrated anecdote involving an Indian scientist from Tamil Nadu: Ramanujan. The great number theorist G.~Hardy recalls~\cite{Hardy}: \textit{I remember once going to see him when he was ill at Putney. I had ridden in taxi cab number 1729 and remarked that the number seemed to me rather a dull one, and that I hoped it was not an unfavourable omen. ``No,'' he replied, ``it is a very interesting number; it is the smallest number expressible as the sum of two cubes in two different ways."}
It is not clear what would Ramanujan had answered if Hardy had taken cab 316. The Penguin Dictionary of Curious and Interesting Numbers~\cite{curious} jumps from 306 to 319 (which cannot be represented as the sum of fewer than 19 4th powers). Even a perusal of \textsl{Wikipedia} does not illuminate, it is a  centred triangular number and a centred heptagonal number, a psalm of John (3:16), the area code of Wichita (Texas), the year of the consulship of Sabinus and Rufinus.

But for Balachandran's student, and collaborators, 316 is very important, it may be said it is a room, but this would be very reductive!
The physical locations is on the third floor of the Syracuse University Physics Building. It is a common room (in the British sense), it has a kitchenette in a corner, a sofa, a table with ensuing chairs, a blackboard. If bibliometrical indices were valid for places rather than people, it would have an h-parameter of at least 40. I will mainly talk of 316 during the eighties, but things did not change in the following decades. The room is still there, there is a photographic proof of it. The blackboard is white, and now there is a projector, and probably the chairs have been reupholstered.

\begin{figure} \centerline{\includegraphics[width=4.5cm]{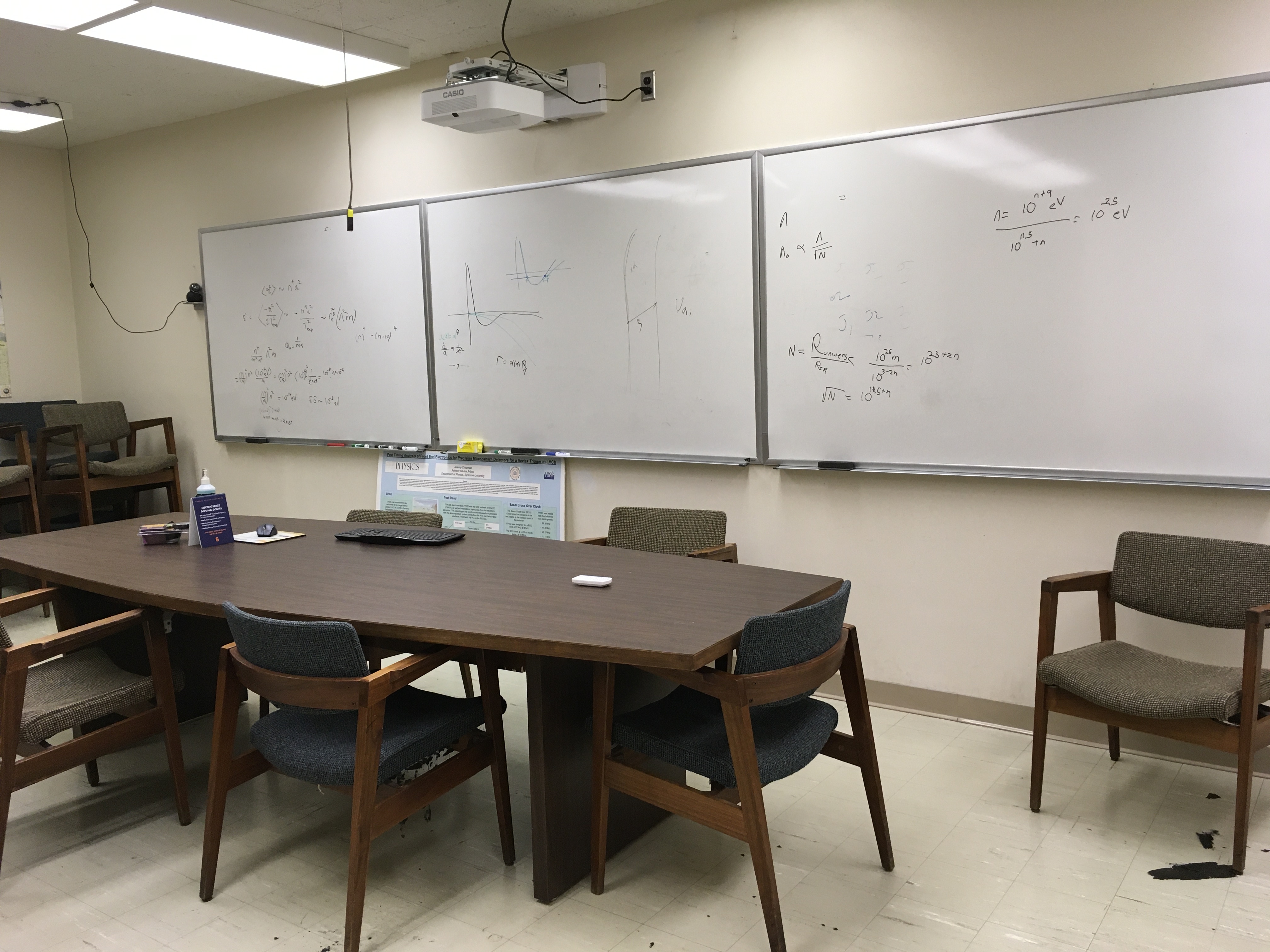}} \caption{Room 316 in 2022.} \label{ra_fig1} \end{figure}

For generations of students and collaborators of Balachandran 316 is not a number, it a locus of the soul. She has nurtured us, in my case gave me some of the best years of my life. There is no doubt in our mind that in that place we became the physicists we are now. More, it shaped us in a more profound way as persons. We met daily in 316, having coffee, discussing physics, solving the problems of the world, gossiping, doing more physics, politics and, you will forgive the vocabulary, but I cannot think of a better word, bullshitting. The room was certainly alive from about twelve noon till about six or seven in the evening, often it was active in the middle of the night. I personally have no idea of its uses before twelve, I was rarely up before the crack of noon. 

Like everything in the universe, 316 is a quantum object, and it has  a ground state. Between about 1~p.m.\  and 5~p.m. the ground state of the room was with Balachandran sitting on the sofa, a student at the board, maybe another more senior person also on the sofa, and other students (or post-doc) at the table, possibly taking notes. The empty room was an excited state, something must have happened (like travelling, or a downtown march against Reagan's politics in Latin America).  

Bal would arrive shortly after lunch. His modus operandi was to wake up early, and work in a room under the roof in his house\footnote{Come to think of it, I have been to Bal's house countless times while I was a student there (1980-1985), enjoyed Indra cuisine, played with Vinod as a child (he is now a surgeon at Sloane-Kettle hospital in New York city), even slept there when I first arrived in Syracuse, but I have no recollection of this other room. The ``room'' for me is only 316.}, then walk along Euclid Avenue to the physics building, attend a few bureaucratic chores, fill a mug with instant coffee and walk to 316. Along the route he would encounter someone, or drop to one of the office and perentoriously say: -get the others!-. And we, the ``al'', of the many papers ``A.P.~Balachandran et al.'', would soon gather to 316. At first we would all be sitting, starting to tell what we had done, a calculation performed, or planned, a paper we had read, a question. Soon we would be explaining to each other all the things we did not understand. Then Bal would address whoever was talking an undescribable gesture with his hand, and the ground state would be reached. The board would be become the center of attention, and the discussion would flow. At times these could be very animated, newcomers had the impression that we were at each other's throat. Other moments there would be long silences, Bal with his chin in his hand, all staring at a blank blackboard, where a wrong idea had been cancelled. Bal would rarely stand up and go himself to the board, when he did it meant that we (but not him) were stuck, or that he had come up with a new important idea.

This little note is too short to write down all of the people who, at one time or another, passed for 316. It suffices to say that in the years I was Bal's student in Syracuse (1980-1985) he has written papers with twenty-one different collaborators. There were remote branches as well, wherever Bal goes there is a 316, in various continents. The very room in which I am presently writing, in my flat, has a whiteboard and an armchair. When I refurbished it I made sure that Bal could have his space during his visits to Napoli. Not far from Napoli the countryside branch, the \emph{Policeta}, near Beppe Marmo's birthplace, has an outdoor room 316, just to mention those nearby. Napoli has several connections with Bal, many of his students were Neapolitans, he visited often and we celebrated his sixtyfifth birthday with a conference in nearby Vietri sul Mare, on the Amalfi coast.

316 survived unscathed Bal's retirement, if you wish to call Bal a retired scientist, I have personally not noticed any difference. The Covid pandemic forced all of us to find new ways to do our work, and 316 found a new life as a virtual seminar room. Bal is running a series of Zoom seminars called 316. So that now 316 can be found also in the cloud!

%

\end{document}